\documentclass[%
 reprint,
 amsmath,amssymb,
 aps,
prb,
showpacs,
longbibliography
]{revtex4-2}

\usepackage{graphicx}
\usepackage{dcolumn}
\usepackage{enumitem}
\usepackage{bm}
\usepackage{physics}
\usepackage{hyperref}
\hypersetup{
    colorlinks,
    linkcolor={blue},
    urlcolor={blue},
    citecolor={blue},
}
\usepackage{color}

\begin{document}

\preprint{APS/123-QED}

\title{Colossal quasiparticle radiation in the Lifshitz spin liquid phase of a \texorpdfstring{\\}{} two-dimensional quantum antiferromagnet}

\author{Matthew C. O'Brien}
\email{matthew.obrien@unsw.edu.au}
\author{Oleg P. Sushkov}
\affiliation{%
School of Physics, The University of New South Wales, Sydney, New South Wales 2052, Australia
}%


\begin{abstract}

Strong quantum fluctuations in magnetic systems can create disordered quantum spin liquid phases of matter which are not predicted by classical physics. The complexity of the exotic phenomena on display in spin liquids has led to a great deal of theoretical and experimental interest. However, understanding the fundamental nature of the excitations in these systems remains challenging. In this work, we consider the Lifshitz quantum critical point in a two-dimensional frustrated $XY$ antiferromagnet. At this point, quantum fluctuations destroy long range order, leading to the formation of an algebraic Lifshitz spin liquid. We demonstrate that the bosonic magnon excitations are long-lived and well-defined in the Lifshitz spin liquid phase, though paradoxically, the dynamic structure factor has a broad non-Lorentzian frequency distribution with no single-particle weight. We resolve this apparent contradiction by showing that the Lifshitz spin liquid suffers from an infrared catastrophe: An external physical probe always excites an infinite number of arbitrarily low energy quasiparticles, which leads to significant radiative broadening of the spectrum.

\end{abstract}

\maketitle 

\section{\label{sec:intro} Introduction}

The nature of excitations in quantum spin liquids --- quantum spin systems with no long range order at $T = 0$ --- is an open question of great importance (see Refs. \cite{wen2004quantum,Savary2017} for reviews), and different models can be host to drastically different excitations. For example, the one-dimensional (1D) antiferromagnetic Heisenberg chain --- which has a disordered ground state --- has spin-1 magnon quasiparticles with a gapped spectrum if the spin per unit cell is an integer \cite{Haldane1983a}. However, if the spin per unit cell is half-integer, then the excitations are gapless spin-1/2 spinons \cite{Faddeev1981}. Further, Kitaev's two-dimensional (2D) honeycomb model is exactly solvable, and has a topological spin liquid phase with spin-1/2 Majorana fermions \cite{Kitaev2006}. In this work, we are interested in disordered states arising from quantum critical fluctuations. In terms of the effect on quasiparticles, two scenarios are known: 1) In the standard Landau-Ginzburg-Wilson (LGW) $O(N)$ universality class description, quasiparticles are not qualitatively changed at the quantum critical point \cite{zinn2002quantum}. 2) Beyond the LGW paradigm, quasiparticles can become fractionalized into ``more elementary'' excitations at deconfined quantum critical points \cite{Senthil2004b}.

The Kitaev model is actually an example of a broader class of systems where the spin liquid ground state is formed by highly frustrated interactions between the microscopic spins \cite{Balents2010}. Classically, magnets with frustrated long range interactions can have collinear and spiral phases which are separated by the so-called Lifshitz point (LP). It is also well understood that quantum fluctuations are always enhanced near the LP \cite{Capriotti2004,Balents2016}. In particular, it has been predicted that in a 2D square lattice antiferromagnet, quantum fluctuations destroy long range order in the vicinity of the LP, leading to the formation of a Lifshitz spin liquid (LSL) \cite{Ioffe1988}. In contrast to other spin liquids, the LSL is connected to the collinear and spiral ordered phases by continuous phase transitions. However, one of us has demonstrated that unlike the $O(3)$ symmetric case, the LSL in an easy-plane $XY$ antiferromagnet only exists at the LP \cite{Kharkov2020}.

In a recent work \cite{OBrien2020}, we demonstrated that significant broadening of the dynamic structure factor can be observed even when the quasiparticles are long-lived and are not fractionalized by strong correlations (eg. in a Luttinger liquid \cite{Schmidt2010}, or at a deconfined critical point \cite{Senthil2004b}). Instead, broadening can be driven by an infrared catastrophe: An infinite number of arbitrarily low energy gapless excitations are emitted by the experimental probe. In the case of the 2D non-critical $XY$ antiferromagnet at finite temperature discussed in that work, the fluctuations were thermal, and essentially classical in nature. However, we noted that the phenomenon of ``soft'' radiation has been well known to particle physicists for decades in the context of the radiation of gauge particles \cite{Bloch1937,Weinberg1965,Greco1975}. There, the phenomenon is driven by the infrared divergence of the virtual processes constituting quantum corrections to scattering cross sections.

In this work, we demonstrate that the LSL in the zero temperature Lifshitz quantum critical $XY$ antiferromagnet is also characterized by an infrared catastrophe. In particular, we illustrate the following: 1) The bosonic quasiparticles are long lived everywhere in the phase diagram. 2) In spite of the narrow quasiparticle linewidth, the dynamic spin structure factor has a broad non-Lorentzian frequency distribution. 3) The LSL has only one effective degree of freedom, in contrast to the $O(N)$ quantum critical point \cite{Chubukov1994}. However, away from the Lifshitz point, this single degree of freedom gives rise to distinct transverse and longitudinal responses.

The rest of this paper is structured as follows: In Sec. \ref{sec:model}, we introduce the square lattice easy-plane $J_1$-$J_3$ model, and the field theory we use to study its low energy sector. This is simply one example of a system capable of realizing Lifshitz criticality. For clarity, we also summarize some general considerations discussed in detail in Ref. \cite{Kharkov2020}. In Sec. \ref{sec:lifetime}, we show that the quasiparticles of the model are long-lived everywhere in the phase diagram, and hence, that their interactions are weak despite the diverging quantum fluctuations near the LP. In Sec. \ref{sec:spectrum}, we derive an expression for the dynamic spin structure factor by resumming diagrams which are formally infinite at the LP. This approach allows us to show that the corrections to the structure factor originate from the incoherent emission of spin waves. Sec. \ref{sec:conclusion} presents our conclusions.


\section{The \texorpdfstring{$J_1$-$J_3$}{J1-J3} Model \label{sec:model}}

A square lattice antiferromagnet with frustrated long range interactions can be described by the $J_1$-$J_3$ model, given by the Hamiltonian
\begin{equation}
    H = J_1 \sum_{\alpha\langle i,j \rangle} S_i^{(\alpha)}S_j^{(\alpha)} + J_3 \sum_{\alpha\langle\langle\langle i,j \rangle\rangle\rangle} S_i^{(\alpha)}S_j^{(\alpha)}, \label{eq:j1j3}
\end{equation}
where $\mathbf{S}_i = (S_i^{(x)}, S_i^{(y)}, S_i^{(z)})$ is the spin operator at site $i$, nested pairs of angle brackets denote summation over nearest, and next-next-nearest neighbor lattice sites, and $J_{1,3} > 0$ is the exchange coupling constant between these pairs of sites, respectively. For an $O(3)$ rotationally symmetric system, summation over spin components is $\alpha = x,y,z$, while for an easy-plane $O(2)$ symmetric system, $\alpha = x,y$ only. We will consider the latter case throughout this paper. The low-lying excitations of this model are bosonic magnons with no definite spin (since the projection of spin is not a good quantum number of the Hamiltonian \eqref{eq:j1j3}). In the N\'eel phase, these correspond physically to in-plane oscillations of the the spins transverse to the direction of staggered magnetization. Hence, quasiparticle states are defined by their momentum $\ket{\mathbf{k}}$. The low energy physics of this lattice model is described by an extended nonlinear sigma model with Lagrangian \cite{Ioffe1988},
\begin{align}
    \mathcal{L} &= \frac{1}{2} \chi_\perp (\partial_t \Vec{n})^2 - \frac{1}{2} \rho (\partial_i \Vec{n})^2 \nonumber \\
    &\qquad- \frac{1}{2} b_1 \Big[ (\partial_x^2 \Vec{n})^2 + (\partial_y^2 \Vec{n})^2 \Big] - b_2 (\partial_x^2 \Vec{n}) (\partial_y^2 \Vec{n}),
\end{align}
where $\Vec{n} = (n_x, n_y)$ is the normalized staggered magnetization $\vec{n}^2 = 1$, $\chi_\perp$ is the magnetic susceptibility in the $\hat{\mathbf{z}}$ direction, $\rho$ is the spin stiffness, and $b_{1,2} \geq 0$ characterize the long range interactions. It is possible to find approximate relations between these parameters and those of the lattice, which we do in Appendix \ref{app:parameters}, but it is well known that the ultraviolet physics of the microscopic model can significantly renormalize these quantities, particularly near quantum critical points. Therefore, we will (mainly) perform calculations in this work without assuming specific values.

When $\rho > 0$ --- which occurs for weak frustration --- the ground state of the system has N\'eel order which spontaneously breaks the rotational symmetry (rotation of $\Vec{n}$, not the coordinate system) of the Lagrangian. When $\rho < 0$ --- which occurs when the long range interactions are strong --- the ground state is spiral ordered: $\varphi(\mathbf{r}) = \mathbf{Q} \cdot \mathbf{r}$, where the direction and magnitude of $\mathbf{Q}$ depend on $b_1$ and $b_2$; the detailed behavior of the system in the spiral phase is beyond the scope of this paper. These two phases are separated classically by the Lifshitz point $\rho = 0$. However, one of us has previously shown that quantum corrections due to the long range interactions lead to a nontrivial renormalization of the spin stiffness $\rho \rightarrow \rho_r$ \cite{Kharkov2020}. For the purposes of this paper, it suffices to note that the LP actually corresponds to $\rho_r = 0$, and from hereon, we will assume that all renormalization has been taken care of and suppress the subscript.

In the $O(2)$ model, it is convenient to parametrize the $\Vec{n}$ field as $\Vec{n} = (\cos\varphi, \sin\varphi)$, and in terms of this angular field, the Lagrangian is
\begin{align}
    \mathcal{L} &= \frac{1}{2} \chi_\perp (\partial_t \varphi)^2 - \frac{1}{2} \rho (\partial_i \varphi)^2 \nonumber \\
    &\qquad - \frac{1}{2} b_1 \Big[ (\partial_x^2 \varphi)^2 + (\partial_y^2 \varphi)^2 \Big] - b_2 (\partial_x^2 \varphi) (\partial_y^2 \varphi) \nonumber \\
    &\qquad - \frac{1}{2} b_1 \Big[ (\partial_x \varphi)^4 + (\partial_y \varphi)^4 \Big] - b_2 (\partial_x \varphi)^2 (\partial_y \varphi)^2. \label{eq:lagrangian_phi}
\end{align}
Therefore, the excitations are bosonic real scalar quasiparticles. Additionally, we see that the frustration has two direct effects on their properties: Firstly, the dispersion is nonlinear in momentum --- due to the terms in the second line of \eqref{eq:lagrangian_phi} --- and in the N\'eel phase, we have
\begin{equation}
    \chi_\perp \omega_{\mathbf{k}}^2 = \rho k^2 + b_1 (k_x^4 + k_y^4) + 2 b_2 k_x^2 k_y^2. \label{eq:dispersion}
\end{equation}
Secondly, the terms in the third line of \eqref{eq:lagrangian_phi} generate interactions between excitations of the $\varphi$ field. The corresponding four point interaction vertex $\mathcal{M}(\mathbf{k}_1,\mathbf{k}_2,\mathbf{k}_3,\mathbf{k}_4)$ is given by the following expression:
\begin{align}
    \mathcal{M} &= \vcenter{\includegraphics[scale=1]{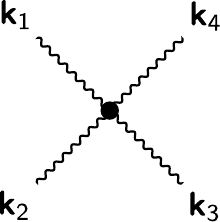}} \nonumber \\
    &= 12 b_1 \Big( k_{1,x} k_{2,x} k_{3,x} k_{4,x} + k_{1,y} k_{2,y} k_{3,y} k_{4,y} \Big) \nonumber \\
    &\ \ + 4 b_2 \Big( k_{1,x} k_{2,x} k_{3,y} k_{4,y} + k_{1,x} k_{2,y} k_{3,x} k_{4,y} \nonumber \\
    &\quad\quad\ \ + k_{1,x} k_{2,y} k_{3,y} k_{4,x} + k_{1,y} k_{2,x} k_{3,x} k_{4,y} \nonumber \\
    &\quad\quad\ \ + k_{1,y} k_{2,y} k_{3,x} k_{4,x} + k_{1,y} k_{2,x} k_{3,y} k_{4,x} \Big). \label{eq:matrix_element}
\end{align}
Therefore, the field theory is not exactly integrable. However, we will demonstrate in the next section that these interactions are very weak.


\section{The Quasiparticle Lifetime \label{sec:lifetime}}

\begin{figure}[!t]
    \centering
    \includegraphics{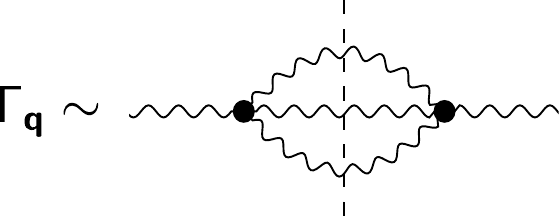}
    \caption{The leading order contribution to the quasiparticle lifetime comes from this two loop diagram.}
    \label{fig:decay_loop}
\end{figure}

As discussed in Ref. \cite{Kharkov2020}, the one loop self energy simply renormalizes the spin stiffness, and otherwise does not have a qualitative effect on the nature of the quasiparticles. At two loop level, the self energy has an imaginary part generated by the diagram in Fig. \ref{fig:decay_loop}, which leads to a finite quasiparticle lifetime. Physically, the frustration allows one quasiparticle to decay spontaneously into three. It is crucial that the frustration leads to a convex dispersion in addition to contributing an interaction vertex, as $1 \rightarrow 3$ decay is forbidden by conservation of energy and momentum in the case of a linear or concave dispersion. The linewidth due to this process is
\begin{align}
    \Gamma_{\mathbf{q}} &= \frac{1}{3!}\frac{1}{(2 \pi)^3 2 \chi_\perp^4 \omega_{\mathbf{q}}} \int \frac{d^2 \mathbf{k}_1}{2\omega_{1}} \frac{d^2 \mathbf{k}_2}{2 \omega_{2}} \frac{d^2 \mathbf{k}_3}{2\omega_{3}} \lvert \mathcal{M}(\mathbf{q}\rightarrow 3\varphi ) \rvert^2 \nonumber \\
    &\times \delta(\omega_{\mathbf{q}} - \omega_1 - \omega_2 - \omega_3) \delta^{(2)}(\mathbf{q} - \mathbf{k}_1 - \mathbf{k}_2 - \mathbf{k}_3), \label{eq:fermi_decay_rate}
\end{align}
where $\mathcal{M}(\mathbf{q}\rightarrow 3\varphi ) = \mathcal{M}(-\mathbf{k}_1,-\mathbf{k}_2,-\mathbf{k}_3,\mathbf{q})$ is the total tree-level probability amplitude for the decay process, as given by \eqref{eq:matrix_element}.

\subsection{Far from the critical point}

\begin{figure}[!b]
    \centering
    \includegraphics[scale=1]{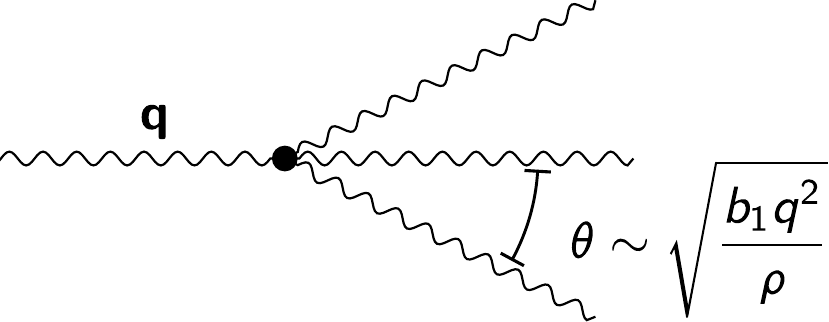}
    \caption{Far from the LP, one quasiparticle can decay into three with momenta lying in a narrow cone centered around the total momentum.}
    \label{fig:jet}
\end{figure}

We begin by considering the system in the N\'eel phase far from the LP. When $b_1 = b_2 = 0$, the dispersion is linear, and conservation of energy and momentum force all three decay products to be emitted in the direction of the initial particle. Hence, if the nonlinearity is very small --- that is, $b_{1,2} q^2 \ll \rho $ --- the decay products are emitted into a very narrow cone centered around the total momentum, as shown in Fig. \ref{fig:jet}. Therefore, if the initial particle is directed along the $\hat{\mathbf{x}}$ direction, the leading order contribution to the decay rate comes only from the very first term in \eqref{eq:matrix_element}. Additionally, the only important nonlinearity in the energy dispersion of each particle comes from the $b_1 k_x^4$ term in \eqref{eq:dispersion}. Using this, we find that the relative linewidth is
\begin{equation}
    \frac{\Gamma_{\mathbf{q}}}{\omega_{\mathbf{q}}} \simeq \frac{1}{(2 \pi)^3} \frac{6 \pi^2}{5005} \left( \frac{b_1 q^2}{\rho} \right)^2 \left( \frac{c q}{\rho} \right)^2 ,
\end{equation}
where $c^2 = \rho/\chi_\perp$ is the wave speed in the linear regime. Not only is $b_1 q^2 \ll \rho$, and the prefactor $\approx 5\times 10^{-5}$, but deep in the ordered N\'eel phase, any low energy excitation will have $c q \ll \rho$ as well. This extremely narrow linewidth $\Gamma_{\mathbf{q}} \lll \omega_{\mathbf{q}}$ is not surprising given that the quasiparticles are non-interacting when frustration is absent. While we expect some qualitative differences in the spiral phase, it is intuitively clear that when the nonlinearity in the dispersion is weak, the quasiparticle lifetime will remain long due to phase space suppression of spontaneous decay.

\subsection{At the Lifshitz point}

The integral \eqref{eq:fermi_decay_rate} is far too impractical to attempt analytically when the nonlinearity of the dispersion is important. Purely from dimensional analysis, we identify that at the LP, the linewidth \eqref{eq:fermi_decay_rate} will have the form
\begin{equation}
    \frac{\Gamma_{\mathbf{q}}}{\omega_{\mathbf{q}}} = \frac{1}{\chi_\perp b_1} \nu(\psi), \label{eq:decay_psi}
\end{equation}
where $\nu(\psi)$ is a number which depends only on the ratio $b_2/b_1$ and the direction of the initial momentum $\hat{\mathbf{k}} = (\cos\psi, \sin\psi)$. We calculate $\nu$ by numerical integration, and present our results in Table \ref{tab:linewidth}. From the lattice parameters \eqref{eq:params}, a naive estimate using the classical LP $J_3 = J_1/4$ gives $\chi_\perp b_1 \approx (3/64) S^2$, which implies the linewidth will be

\renewcommand{\arraystretch}{1.5}
\begin{table}[!b]
    \caption{The numerical coefficient of the linewidth $\nu(\psi)$, for different ratios of frustration constants $b_2/b_1$. The mean value presented is an angular average on the interval $(0,\pi/2)$. Meaningful (larger than the numerical uncertainty) angular variation in $\nu$ was only found for $b_2 = 0$.}
    \begin{ruledtabular}
    \begin{tabular}{ccc}
        $b_2/b_1$ & $\mathrm{Mean}[\nu(\psi)]$ & Range$[\nu(\psi)]$ \\ \colrule
        0 & $5.1 \times 10^{-3}$ & $0.3 \times 10^{-3}$   \\
        0.5 & $3.2 \times 10^{-3}$ & --- \\
        1 & $2.6 \times 10^{-3}$ & --- \\
    \end{tabular}
    \end{ruledtabular}
    \label{tab:linewidth}
\end{table}

\begin{equation}
    \Gamma_{\mathbf{q}}\sim 0.1 S^{-2} \omega_{\mathbf{q}}. \label{eq:linewidth_lifshitz}
\end{equation}
Therefore, the linewidth will be small $\Gamma_{\mathbf{q}} \lesssim \omega_{\mathbf{q}}$, even for $S = 1/2$. As noted previously, there will be quantum corrections to the field theory parameters, so we might expect the coefficient of \eqref{eq:linewidth_lifshitz} to be different by a factor of 2. However, this does not change the result qualitatively, since the relative linewidth is simply a constant which is never parametrically large.

We also note the following: 1) The integral \eqref{eq:fermi_decay_rate} is very well convergent in the infrared due to the momentum dependence of the matrix element \eqref{eq:matrix_element}. In fact, the dominant contribution comes from when the momentum is evenly distributed among the decay products. 2) Spiral order only affects the linear part of the quasiparticle dispersion \cite{Kharkov2020}. 3) If $b_{1,2} q^2 \gg \rho$, then the linear terms in the dispersion are irrelevant. Therefore, \eqref{eq:decay_psi} and the numerical results for $\nu(\psi)$ give the leading order contribution to the decay rate on \textit{both} sides of the LP.

In this section, we have demonstrated that the quasiparticle linewidth (its inverse lifetime) is small compared to its energy everywhere in the phase diagram, including in the Lifshitz spin liquid phase. This indicates that the nature of the quasiparticles in the LSL is not qualitatively different from in the ordered phases.



\section{Spectral Broadening Near the Lifshitz Point \label{sec:spectrum}}

\begin{figure*}
    \centering
    \includegraphics[scale=0.85]{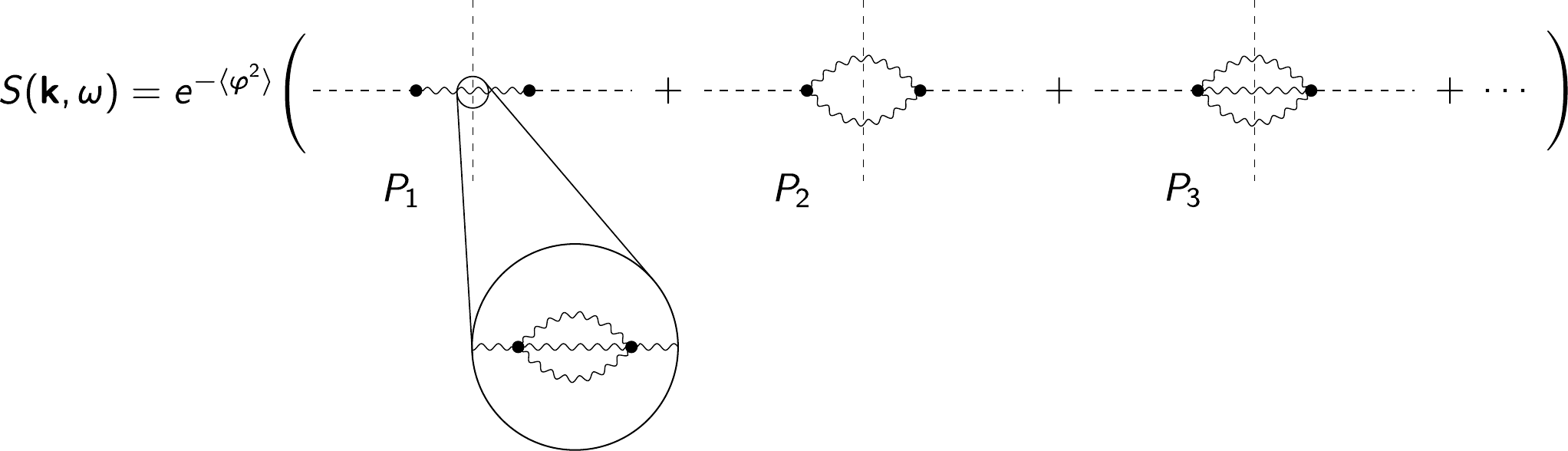}
    \caption{Diagrammatic expansion of the dynamic structure factor. Dashed lines denote the source, wavy lines denote quasiparticles, and vertical dashing denotes application of cut rules. The ``magnifying'' inset shows higher order off-shell processes which generate broadening.
    }
    \label{fig:diagram_exp}
\end{figure*}

The dynamic structure factor is directly measurable through inelastic neutron scattering, which is typically the experiment of choice for studying magnetic systems, including spin liquids \cite{Banerjee2017}:
\begin{equation}
    S_{ij}(\mathbf{k},\omega) = \int d t\, d^2 \mathbf{r}\, \bra{0} n_i(\mathbf{r},t) n_j(0) \ket{0}\, e^{i(\omega t - \mathbf{k}\cdot \mathbf{r})}. \label{eq:skw}
\end{equation}
This expression would, for example, correspond to the response from a polarized neutron source. Following the approach of Ref. \cite{OBrien2020}, we expand the structure factor in a spectral representation \cite{Lifshitz1995},
\begin{align}
    S_{ij}(\mathbf{k},\omega) &= \sum_{\alpha} \lvert \bra{\alpha} n_i (0) \ket{0} \rvert^2 \nonumber \\ &\qquad\quad \times(2\pi)^3 \delta_{ij}  \delta(\omega - \omega_\alpha) \delta^{(2)}(\mathbf{k} - \mathbf{k}_\alpha), \label{eq:skw_spectral}
\end{align}
where $\ket{\alpha}$ is an excited $\varphi$ quasiparticle state, and $\omega_\alpha$ and $\mathbf{k}_\alpha$ are the energy and momentum of that state. We fix the direction of the spontaneous staggered magnetization to correspond to $\varphi = 0$, and hence, and if $\ket{\alpha} = \ket{\mathbf{k}_1,\dots,\mathbf{k}_m}$ is an $m$-particle excited state, the matrix element in \eqref{eq:skw_spectral} is
\begin{equation}
    \lvert\bra{\alpha} n_i(0) \ket{0}\rvert^2 = e^{-\langle\varphi^2 \rangle} \prod_{j=1}^m \frac{1}{2 \chi_\perp \omega_{\mathbf{k}_j}}, \label{eq:skw_matrix}
\end{equation}
for an even (odd) number of particles if $i = x$ ($y$), and zero otherwise. Therefore, when the external source interacts with the system, it couples to the quantum fluctuations of the $\varphi$ field
\begin{equation}
    \langle\varphi^2 \rangle = \int_0^\Lambda \frac{d^2 \mathbf{q}}{(2 \pi)^2} \frac{1}{2 \chi_\perp \omega_{\mathbf{q}}} \simeq \frac{\kappa}{8 \pi \sqrt{\chi_\perp b_1}} \log \frac{4 b_1 \Lambda^2}{\rho}, \label{eq:phi_fluct}
\end{equation}
assuming $b_{1,2} \Lambda^2 \gg \rho$, and where $\Lambda \sim \pi/a$ is the ultraviolet momentum cut-off, $a$ is the microscopic lattice spacing, and
\begin{equation}
    \kappa = \int_0^{2\pi} \frac{d\psi/(2\pi)}{\sqrt{\cos^4\psi + \sin^4\psi + 2 (b_2/b_1) \cos^2\psi \sin^2\psi}}.
\end{equation}
From this, we see that processes involving a finite number of quasiparticles are suppressed in the limit $\rho \rightarrow 0$, becoming forbidden precisely at the LP. Therefore, to obtain an expression for the structure factor which describes the Lifshitz spin liquid, it is necessary to sum the spectral representation \eqref{eq:skw_spectral} to all orders in the number of emitted quasiparticles. For future use, we note here that the form of \eqref{eq:phi_fluct} indicates that the smallest physically important quasiparticle momentum is 
\begin{equation}
    q_{\mathrm{min}} = \sqrt{\frac{\rho}{4 b_1}}. \label{eq:qmin}
\end{equation}
That is, in close proximity to the LP --- where $\rho$ is the smallest energy scale --- it suffices to set $\rho = 0$ in all calculations as long as any infrared divergent integrals are cut-off by $q_{\mathrm{min}}$.


Just as in Ref. \cite{OBrien2020}, we can define the integrated probability for the source to excite $m$ quasiparticles given a transfer of energy $\omega$ and momentum $\mathbf{k}$
\begin{align}
    P_m(\mathbf{k},\omega) &= \frac{1}{m!}\int d \lbrace\mathbf{k_\alpha}\rbrace\, \lvert \bra{\mathbf{k}_1,\dots,\mathbf{k}_m}\Vec{n}(0) \ket{0} \rvert^2 \nonumber \\
    &\qquad\qquad \times (2 \pi)^3 \delta(\omega - \omega_\alpha) \delta^{(2)}(\mathbf{k} - \mathbf{k}_\alpha), \label{eq:pm_int}
\end{align}
where $d \lbrace\mathbf{k_\alpha}\rbrace = \prod_{j=1}^m d^2 \mathbf{k}_j/(2 \pi)^2$, so that
\begin{equation}
    S_{ii}(\mathbf{k},\omega) = \sum_{m} P_m(\mathbf{k},\omega), \label{eq:skw_sum}
\end{equation}
summing over even $m$ if $i = x$ and odd $m$ if $i = y$. This defines a diagrammatic expansion which is shown in Fig. \ref{fig:diagram_exp}. As expected, the elastic peak corresponding to the static order has a spectral weight which is reduced by proximity to the spin liquid phase:
\begin{equation}
    P_0(\mathbf{k},\omega) = e^{-\langle\varphi^2\rangle}(2\pi)^3 \delta(\omega) \delta^{(2)}(\mathbf{k}). \label{eq:static_peak}
\end{equation}
However, \eqref{eq:skw_sum} does not actually account for all possible processes. Since the quasiparticle lifetime is finite, the probe can also excite quasiparticles which are off-shell, which then decay into three on-shell particles, as shown in the inset to Fig. \ref{fig:diagram_exp}. Therefore, in \eqref{eq:pm_int}, we replace the energy $\delta$ function by a Lorentzian distribution:
\begin{equation}
    (2 \pi) \delta(\omega - \omega_\alpha) \longrightarrow \frac{\Gamma_\alpha}{(\omega - \omega_\alpha)^2 + \Gamma_\alpha^2/4},
\end{equation}
where $\Gamma_\alpha$ is the total width associated with the off-shell process. For example, the single particle $\delta$ peak becomes broadened:
\begin{equation}
    P_1(\mathbf{k},\omega) = e^{-\langle\varphi^2\rangle} \frac{1}{2 \chi_\perp \omega_{\mathbf{k}}} \frac{\Gamma_{\mathbf{k}}}{(\omega - \omega_{\mathbf{k}})^2 + \Gamma_{\mathbf{k}}^2/4}, \label{eq:single_particle_peak}
\end{equation}
where $\Gamma_{\mathbf{k}}$ is the single particle inverse lifetime determined in Sec. \ref{sec:lifetime}. Therefore, by broadening the energy conserving $\delta$ function, our definition of $P_m$ corresponds to the probability of exciting $m$ quasiparticles, some of which may be off-shell. However, we can see from \eqref{eq:single_particle_peak} that if $\lvert \omega - \omega_{\mathbf{k}} \rvert \doteq \lvert \Delta \rvert \gg \Gamma_{\mathbf{k}}$, the peak has zero effective width. This defines two regimes of interest: $\lvert \Delta \rvert \gg \Gamma_{\mathbf{k}}$, where we can neglect the contribution from off-shell processes to the structure factor and retain the energy $\delta$ function in \eqref{eq:pm_int}, and $\lvert \Delta \rvert \ll \Gamma_{\mathbf{k}}$, where we must take the lifetime into account.

\subsection{Regime 1: Away from resonance}

Away from the resonance $\omega = \omega_{\mathbf{k}}$, the single particle peak has no influence, so we consider the two particle diagram shown in Fig. \ref{fig:diagram_exp},
\begin{align}
    P_2(\mathbf{k},\omega) &= \frac{e^{-\langle\varphi^2\rangle}}{2!} \int \frac{d^2 \mathbf{q}}{(2 \pi)^2 2 \chi_\perp \omega_{\mathbf{q}}} \frac{1}{2 \chi_\perp \omega_{\mathbf{k-q}}} \nonumber \\
    &\qquad\qquad\qquad \times (2 \pi) \delta(\omega - \omega_{\mathbf{q}} - \omega_{\mathbf{k-q}}),
\end{align}
which has no infrared divergence, but has a singularity at $\mathbf{q} = \mathbf{k}$ when $\omega = \omega_{\mathbf{k}}$. Therefore, in the regime $ \Gamma_{\mathbf{k}} \ll \lvert\Delta\rvert \ll \omega_{\mathbf{k}}$, most of the momentum from the source will be transferred to a single quasiparticle; when $b_1 = b_2$, the integral can be evaluated exactly to confirm this inference. For general frustration, we find in the limit $\Gamma_{\mathbf{k}} \ll \lvert\Delta\rvert \ll \omega_{\mathbf{k}}$ that
\begin{equation}
    P_2(\mathbf{k},\omega) \simeq \frac{\kappa e^{-\langle\varphi^2\rangle}}{8 \sqrt{\chi_\perp^3 b_1} \omega_{\mathbf{k}} \lvert \omega - \omega_{\mathbf{k}} \rvert}.
\end{equation}
It is interesting to compare this result with the case of the non-frustrated $XY$ model discussed in Ref. \cite{OBrien2020}: In that case, anti-Stokes ($\Delta < 0$) processes were forbidden at zero temperature by the phase space constraints imposed by a linear dispersion. At finite temperature, they became allowed if the external probe absorbed at least one excitation from the thermal bath. In our present work, the quadratic dispersion near the LP allows conservation of energy and momentum to be satisfied for positive and negative detuning $\Delta$. However, $\omega < 0$ processes remain forbidden since it is impossible to extract energy from the quantum vacuum.

The probability of exciting three particles can be evaluated within the same approximation. However, since the quasiparticle dispersion is not isotropic if $b_1 \neq b_2$ ($\omega_{\mathbf{k}}$ depends on the direction of $\mathbf{k}$) an accurate separation of momentum scales between the two lower energy excitations should take this into account. Linearizing the energy $\delta$ constraint with respect to the small momenta $\mathbf{q}$ and $\mathbf{p}$,
\begin{equation}
    \omega - \omega_{\mathbf{q}} - \omega_{\mathbf{p}} - \omega_{\mathbf{k-q-p}} \simeq \Delta + \frac{\partial \omega_{\mathbf{k}}}{\partial \mathbf{k}} \cdot (\mathbf{q} + \mathbf{p}),
\end{equation}
imposes the limits of integration
\begin{equation}
    q_{\mathrm{min}} < q < \frac{\lvert \Delta \rvert}{v_{\mathbf{k}}}, \qquad \text{ and } \qquad \frac{\lvert \Delta \rvert}{v_{\mathbf{k}}} < p < \infty,
\end{equation}
where we extend the domain of $p$ to $\infty$ since the contribution at large momenta is negligible, and 
\begin{equation}
    v_{\mathbf{k}} = \left\lvert \frac{\partial \omega_{\mathbf{k}}}{\partial \mathbf{k}} \right\rvert,
\end{equation}
is the group velocity. Given this, we find that
\begin{align}
    P_3(\mathbf{k},\omega) &\simeq P_2(\mathbf{k},\omega) \times \int_{q_{\mathrm{min}}}^{\lvert\Delta\rvert/v_{\mathbf{k}}} \frac{d^2 \mathbf{q}}{(2 \pi)^2} \frac{1}{2 \chi_\perp \omega_{\mathbf{q}}} \nonumber \\
    &= P_2(\mathbf{k},\omega) \times \frac{\kappa}{4 \pi \sqrt{\chi_\perp b_1}} \log \left( \frac{\lvert \Delta \rvert}{v_{\mathbf{k}} q_{\mathrm{min}}} \right).
\end{align}
Within the regime we are working in, as shown in Ref. \cite{OBrien2020}, this factorization of the emission probability holds to all orders in the particle number. That is, for a total of $N$ excitations, there will be $N - 2$ ``soft'' quasiparticles emitted in a statistically independent manner with a uniform angular distribution (for fixed momentum), and
\begin{align}
    P_N(\mathbf{k},\omega) &\simeq \frac{1}{(N - 2)!}  P_2(\mathbf{k},\omega) \nonumber \\
    &\quad \times \left[ \frac{\kappa}{4 \pi \sqrt{\chi_\perp b_1}} \log \left( \frac{\lvert \Delta \rvert}{v_{\mathbf{k}} q_{\mathrm{min}}} \right) \right]^{N-2}. \label{eq:p_n}
\end{align}
We then insert this expression into \eqref{eq:skw_sum} to find that
\begin{equation}
    S_{ii}(\mathbf{k},\omega) = P_2(\mathbf{k},\omega) \times \begin{cases} \cosh \lambda_1 & \text{for } i = x, \\ \sinh \lambda_1 & \text{for } i = y,  \end{cases}
\end{equation}
where
\begin{equation}
    \lambda_1 =  \frac{\kappa}{4 \pi \sqrt{\chi_\perp b_1}} \log \left( \frac{\lvert \Delta \rvert}{v_{\mathbf{k}} q_{\mathrm{min}}} \right), \label{eq:mean_emission}
\end{equation}
is the mean number of excited quasiparticles with momentum less than $\lvert\Delta\rvert/v_{\mathbf{k}}$. It is straightforward to see that the quantum fluctuations in the vertex $e^{-\langle\varphi^2\rangle}$ \eqref{eq:phi_fluct} regularize the infrared divergence of $X$ in the limit $\rho \rightarrow 0$, and that in the LSL, we have
\begin{align}
    &S_{xx}(\mathbf{k},\omega) = S_{yy}(\mathbf{k},\omega) \nonumber \\
    &\qquad = \frac{\kappa }{16 \sqrt{\chi_\perp^3 b_1} \omega_{\mathbf{k}} \lvert \omega - \omega_{\mathbf{k}} \rvert} \left( \frac{\lvert \omega - \omega_{\mathbf{k}} \rvert}{\Lambda v_{\mathbf{k}}} \right)^{\zeta}, \label{eq:skw_xx_yy}
\end{align}
where
\begin{equation}
    \zeta = \frac{\kappa}{4 \pi \sqrt{\chi_\perp b_1}},
\end{equation}
is the same critical exponent found in Ref. \cite{Kharkov2020}.

\begin{figure}[!b]
    \centering
    \includegraphics[scale=0.9]{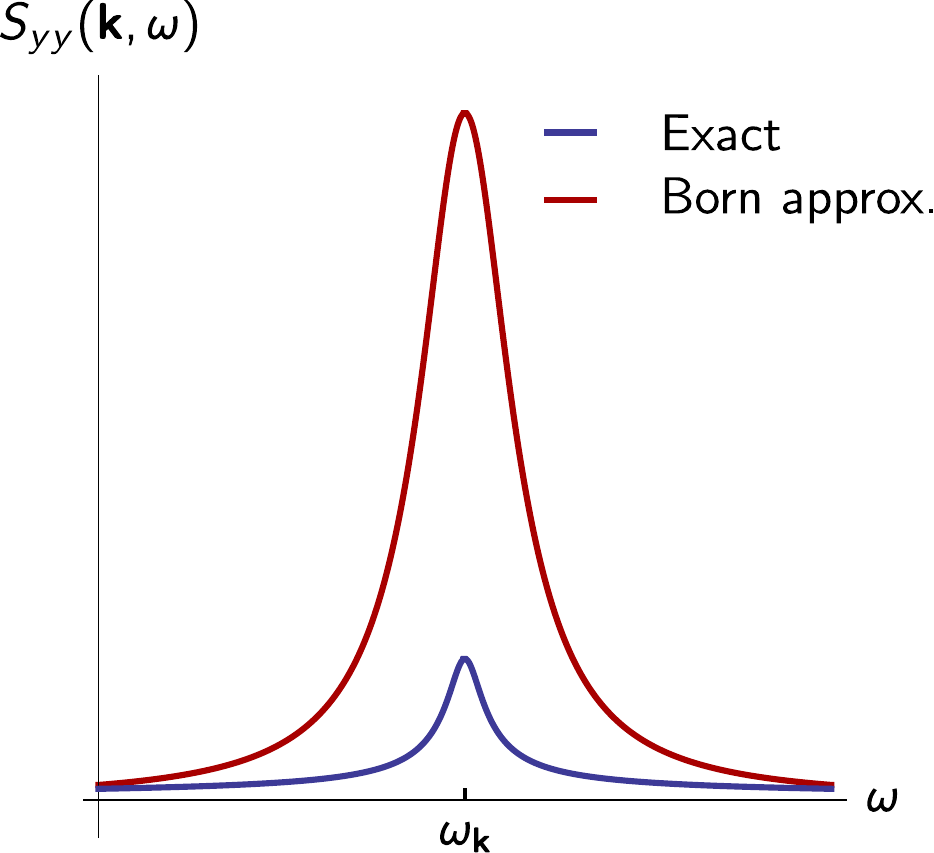}
    \caption{The dynamic structure factor in the LSL phase, comparing the exact solution \eqref{eq:skw_xx_yy} and \eqref{eq:skw_gamma} with a naive Born approximation. Parameters: $\sqrt{b_1} k^2 = 1$, $\sqrt{b_1}\Lambda^2 = 10$, $\chi_\perp = 1$, $\zeta = 0.2$, and $\Gamma_{\mathbf{k}} = 0.3$.}
    \label{fig:spectrum}
\end{figure}

\subsection{Regime 2: On Resonance}

To avoid cumbersome equations, we will only consider the case $\Delta = 0$; this will reveal all of the important physics. In this case, 
\begin{equation}
    P_1(\mathbf{k}, \omega_{\mathbf{k}}) = \frac{2 e^{-\langle\varphi^2\rangle}}{\chi_\perp \omega_{\mathbf{k}} \Gamma_{\mathbf{k}}},
\end{equation}
and the probability of exciting two particles is
\begin{align}
    P_2(\mathbf{k}, \omega_{\mathbf{k}}) &\simeq \frac{\kappa e^{-\langle\varphi^2\rangle}}{4 \sqrt{\chi_\perp^3 b_1} \omega_{\mathbf{k}}} \nonumber \\
    &\qquad \times \int \frac{d^2 \mathbf{q}}{(2 \pi)^2} \frac{1}{q^2} \frac{\Gamma}{(v_{\mathbf{k}} q \cos\theta_{\mathbf{q}})^2 + \Gamma^2/4},
\end{align}
which is now infrared divergent. In the regime $\lvert\Delta\rvert \gg \Gamma_{\mathbf{k}}$, we found that both of the particles in this process would have a finite energy, with one particle carrying $\sim \lvert\Delta\rvert$. By reducing the detuning, we also reduce the energy of this second particle until it is cut off by $v_{\mathbf{k}} q_{\mathrm{min}}$. Consequently, the broadening from the soft quasiparticle is negligible, and we approximate the total width of the two particle process as $\Gamma \approx \Gamma_{\mathbf{k}}$, independent of $\mathbf{q}$. This leads to
\begin{align}
    P_2(\mathbf{k}, \omega_{\mathbf{k}}) &\simeq P_1(\mathbf{k}, \omega_{\mathbf{k}}) \times \int_{q_{\mathrm{min}}}^{\Gamma_{\mathbf{k}}/v_{\mathbf{k}}} \frac{d^2 \mathbf{q}}{(2 \pi)^2} \frac{1}{2 \chi_\perp \omega_{\mathbf{q}}} \nonumber \\
    &= P_1(\mathbf{k}, \omega_{\mathbf{k}}) \times  \frac{\kappa}{4 \pi \sqrt{\chi_\perp b_1}} \log \left( \frac{\Gamma_{\mathbf{k}}}{v_{\mathbf{k}} q_{\mathrm{min}}} \right).
\end{align}
Therefore, we find very similar factorization of the probability, and it is immediately clear that
\begin{equation}
    P_N(\mathbf{k}, \omega_{\mathbf{k}}) = \frac{1}{(N - 1)!} P_1(\mathbf{k}, \omega_{\mathbf{k}}) \lambda_2^{N-1},
\end{equation}
where
\begin{equation}
    \lambda_2 = \frac{\kappa}{4 \pi \sqrt{\chi_\perp b_1}} \log \left( \frac{\Gamma_{\mathbf{k}}}{v_{\mathbf{k}} q_{\mathrm{min}}} \right), \label{eq:mean_emission_r2}
\end{equation}
is the mean number of emitted particles when the source is on resonance. Then, in the LSL phase, summing up all $N$ gives the structure factor
\begin{equation}
    S_{xx}(\mathbf{k},\omega_{\mathbf{k}}) = S_{yy}(\mathbf{k},\omega_{\mathbf{k}}) = \frac{1}{\chi_\perp \omega_{\mathbf{k}} \Gamma_{\mathbf{k}}} \left( \frac{\Gamma_{\mathbf{k}}}{\Lambda v_{\mathbf{k}}} \right)^{\zeta}. \label{eq:skw_gamma}
\end{equation}
This is a strikingly different expression to \eqref{eq:skw_xx_yy}. Naively, we might have expected that crossing over to the regime $\lvert\Delta\rvert \ll \Gamma_{\mathbf{k}}$ would have the effect of replacing
\begin{equation}
    \lvert \omega - \omega_{\mathbf{k}} \rvert \xrightarrow{\text{wrong}} \Gamma_{\mathbf{k}}.
\end{equation}
Instead, we we actually find
\begin{equation}
    \lvert \omega - \omega_{\mathbf{k}} \rvert \xrightarrow{\text{correct}} \frac{\kappa}{16 \sqrt{\chi_\perp b_1}} \Gamma_{\mathbf{k}}.
\end{equation}
This reflects the change from a spectrum characterized by the emission of two particles in a range of possible directions, to a predominantly single particle spectrum (with radiative corrections) which is characterized by the finite lifetime of the excitations. 

As shown in Fig. \ref{fig:spectrum}, the finite quasiparticle lifetime regulates the singularity in the spectrum at $\omega = \omega_{\mathbf{k}}$. This is in contrast to the non-critical $XY$ antiferromagnet at finite temperature discussed in Ref. \cite{OBrien2020}, where the quasiparticles were non-interacting and infinitely long-lived; the spectrum contained a physical singularity. For comparison, we also plot the naive Born approximation for the transverse component of the structure factor, which corresponds to the diagram $P_1$ shown in Fig. \ref{fig:diagram_exp}, \textit{without} the vertex $e^{-\langle\varphi^2\rangle}$. Evidently, this first order approximation differs considerably from the exact solution obtained by summing diagrams to all orders.

\subsection{Discussion}

From the results obtained, we can deduce the following key points:

\begin{enumerate}[label=(\arabic*),wide=0pt,labelindent=\parindent]
    \item Lifshitz criticality leads to remarkably similar physics to the effect of temperature in the non-frustrated $XY$ model: The quantum fluctuations ``dress'' the probe so that interactions with real and virtual particles become indistinguishable. In the spin liquid phase, the probability of exciting any finite number of quasiparticles is forced to zero by the vertex $e^{-\langle\varphi^2\rangle}$, and so the source must excite an infinite number of arbitrarily low energy excitations.
    \item Just like the vanishing elastic spectral weight at the LP, the absence of polarization dependence --- the structure factor has $S_{xx} = S_{yy}$ --- is a key signature of the absence of long range order in the LSL. 
    \item The lattice parameters \eqref{eq:params} imply that the critical exponent $\zeta \sim 1/S$. This indicates that the radiative corrections are largest for smaller values of spin per unit cell $S$. Additionally, the expressions for the structure factor \eqref{eq:skw_xx_yy} and \eqref{eq:skw_gamma} account for all orders in $1/S$. In this sense, our method reveals physics which cannot be captured by linear spin wave theory.
    \item Given the definition of the mean number of emitted quasiparticles \eqref{eq:mean_emission}, it is simple to show that the average intensity of the soft radiation, integrated over the direction of emission, is independent of frequency:
    \begin{equation}
        d I = \omega d X = \frac{\kappa}{8 \pi \sqrt{\chi_\perp b_1}}  d \omega.
    \end{equation}
    This mirrors the behavior of \textit{bremsstrahlung} in quantum electrodynamics (QED), where the emitted photons have an energy distribution which is equal to the frequency-independent classical expression for the radiated energy \cite{Berestetskii1982}. In the non-relativistic limit, the intensity is \cite{Landau1987},
    \begin{equation}
        d I_{\text{QED}} = \frac{2 e^2 v^2}{3\pi c^3}  d\omega,
    \end{equation}
    where $v$ is the speed of the radiating charged particle. By analogy, we consider that the radiation of spin waves in a LSL is ``as classical as possible''.
    \item The number of effective degrees of freedom can be quantified by the specific heat capacity as $T \rightarrow 0$; this counts the number of gapless modes. Since the quasiparticles remain well-defined everywhere in the phase diagram, the specific heat of the LSL will be the same as that of a single quadratic degree of freedom. This is in contrast with $O(N)$ quantum criticality: In the broken symmetry phase, longitudinal excitations of the order parameter are gapped out, and there are $N - 1$ independent gapless linear Goldstone modes. In the quantum disordered phase, there are $N$ independent gapped modes. However, while this gap vanishes at the $O(N)$ critical point, the number of effective degrees of freedom is a rational number $4 N /5$ \cite{Chubukov1994,Shevchenko2000}. 
    \item The problem we have studied in this work is a zero temperature unitary quantum theory. Therefore, the cancellation of infrared divergences in physical observables such as the dynamic structure factor is guaranteed by the KLN theorem \cite{Kinoshita1962,Lee1964}.
\end{enumerate}

\subsection{Verification of sum rules}

To confirm the validity of our result for the dynamic structure factor \eqref{eq:skw_xx_yy}, we calculate its static sum rule \cite{Lifshitz1995}, using two different approaches. Defining the total unpolarized response
\begin{align}
    S(\mathbf{k},\omega) &= S_{xx}(\mathbf{k},\omega) + S_{yy}(\mathbf{k},\omega) \nonumber \\
    &= \frac{\kappa }{8 \sqrt{\chi_\perp^3 b_1} \omega_{\mathbf{k}} \lvert \omega - \omega_{\mathbf{k}} \rvert} \left( \frac{\lvert \omega - \omega_{\mathbf{k}} \rvert}{\Lambda v_{\mathbf{k}}} \right)^{\zeta}, \label{eq:skw_unpolarized}
\end{align}
the static structure factor is
\begin{align}
    S(\mathbf{k}) &= \int \frac{d \omega}{2 \pi} S(\mathbf{k},\omega)\nonumber \\
    &= \int d^2 \mathbf{r}\, \langle \Vec{n}(\mathbf{r},0)\cdot\Vec{n}(0) \rangle\, e^{-i \mathbf{k}\cdot\mathbf{r}}, \label{eq:static}
\end{align}
where the equal time correlator in the LSL phase was calculated in Ref. \cite{Kharkov2020}, for $\Lambda r \gg 1$,
\begin{equation}
    \langle \Vec{n}(\mathbf{r},0)\cdot\Vec{n}(0) \rangle \simeq (\Lambda r)^{-\zeta}. \label{eq:correlator}
\end{equation}
Inserting this expression into \eqref{eq:static} and performing the Fourier transform, we obtain
\begin{equation}
    S(\mathbf{k}) \simeq \frac{\kappa}{2 \sqrt{\chi_\perp b_1} k^2} \left( \frac{k}{\Lambda} \right)^{\zeta}. \label{eq:static_fourier}
\end{equation}
We then compare this with our result obtained from summing the spectral representation: The spectrum is sharply peaked at $\omega_{\mathbf{k}}$, and the singularity at this point in \eqref{eq:skw_unpolarized} is integrable. Therefore, we can neglect the effect of the lifetime near resonance, and the $\omega$ integral in \eqref{eq:static} is well approximated by a restriction to the interval $(0, 2\omega_{\mathbf{k}})$. This yields 
\begin{equation}
    S(\mathbf{k}) \simeq \frac{1}{2 \chi_\perp \omega_{\mathbf{k}}} \left( \frac{\omega_{\mathbf{k}}}{\Lambda v_{\mathbf{k}}} \right)^\zeta, \label{eq:static_v2_0}
\end{equation}
which initially seems quite different to \eqref{eq:static_fourier}. However, the radiation which contributes the correction $(\omega_{\mathbf{k}}/\Lambda v_{\mathbf{k}})^\zeta$ has an extremely long wavelength, and hence will only weakly ``feel'' any anisotropy. It is also emitted statistically independently in all directions. Therefore, inside the parentheses, we can safely set $b_2 = b_1$, and then average \eqref{eq:static_v2_0} over the direction of $\mathbf{k}$ (this gives the correct factor of $\kappa$). After doing so, we obtain precisely \eqref{eq:static}. We also observe that
\begin{equation}
    \int_0^\Lambda \frac{d^2 \mathbf{k}}{(2 \pi)^2} S(\mathbf{k}) = 1,
\end{equation}
which is exactly equal to the elastic spectral intensity far away from the LP (see \eqref{eq:static_peak}). Therefore, just like temperature in the non-critical $XY$ model, Lifshitz criticality ``disolves'' the elastic peak into an infinite particle spectrum.

\subsection{Away from the Lifshitz point}

We conclude this section by examining the properties of the system slightly away from the LP. Firstly, we note that the statistically independent emission of the soft radiation --- which is a consequence of the factorization of the probability $P_N$ --- implies that the number of emitted low energy quasiparticles will be Poisson distributed:
\begin{equation}
    P(\text{emit }N\varphi) = \frac{1}{N!} \lambda^N e^{-\lambda}, \label{eq:poisson}
\end{equation}
where $\lambda$ is the mean number of soft excitations, given by either \eqref{eq:mean_emission} or \eqref{eq:mean_emission_r2}, depending on the frequency regime, as described above. As a result, $\lambda$ will also be the variance in the number of soft excitations, and diverges as $\rho \rightarrow 0$. However, the number of particles emitted in any range of momenta $0 < k_{-} < k_{+} \ll \mathrm{max}\lbrace\lvert\Delta\rvert, \Gamma_{\mathbf{k}}\rbrace /v_{\mathbf{k}}$ remains finite at the LP, and the distribution is obtained by replacing $\lambda$ in \eqref{eq:poisson} with
\begin{equation}
    \bar{n} = \frac{\kappa}{4 \pi \sqrt{\chi_\perp b_1}} \log \frac{k_{+}}{k_{-}}.
\end{equation}

Secondly, the dynamic structure factor \eqref{eq:skw_xx_yy} is polarization dependent away from the LP. From a fundamental perspective, it is interesting that a single well-defined quasiparticle degree of freedom gives rise to an anisotropic polarization response. This dependence is due to the nontrivial manner in which the source couples to the excitations via the cosine or sine of the angle $\varphi$: Away from the LP, only an even (odd) number of quasiparticles are excited if the source is polarized parallel (perpendicular) to the N\'eel order parameter. We also calculate the $\omega$ integrated spectral weight contained in each polarization channel. Close to the LP, the difference between the tails of the two polarizations is
\begin{align}
    &S_{xx}(\mathbf{k},\omega) - S_{yy}(\mathbf{k},\omega) \nonumber \\
    &\qquad = \frac{\kappa }{8 \sqrt{\chi_\perp^3 b_1} \omega_{\mathbf{k}} \lvert \omega - \omega_{\mathbf{k}} \rvert} \left( \frac{q_{\mathrm{min}}^2 v_{\mathbf{k}}}{\Lambda \lvert \omega - \omega_{\mathbf{k}}  \rvert} \right)^\zeta.
\end{align}
Since we derived this expression in the limit $\lvert \Delta \rvert/v_{\mathbf{k}} \gg q_{\mathrm{min}}$, the difference between the spectral weights of each channel will be
\begin{align}
    &\simeq 2 \int_{v_{\mathbf{k}} q_{\mathrm{min}}}^\infty \frac{d \Delta}{2 \pi} \left[ S_{xx}(\mathbf{k},\omega) - S_{yy}(\mathbf{k},\omega) \right] \nonumber \\
    &= \frac{1}{2 \chi_\perp \omega_{\mathbf{k}}} \left( \frac{q_{\mathrm{min}}}{\Lambda} \right)^\zeta, \label{eq:weight_diff}
\end{align}
where $q_{\mathrm{min}} = \sqrt{\rho/4 b_1}$. However, away from the LP, $S_{yy}(\mathbf{k},\omega)$ also contains a finite single particle $\delta$ peak \eqref{eq:single_particle_peak},
\begin{equation}
    P_1(\mathbf{k},\omega) \simeq \frac{1}{2 \chi_\perp \omega_{\mathbf{k}}} \left( \frac{q_{\mathrm{min}}}{\Lambda} \right)^\zeta \delta(\omega - \omega_{\mathbf{k}}),
\end{equation}
which has an integrated weight exactly equal to \eqref{eq:weight_diff}. Therefore, the \textit{total} integrated spectral weights of the two channels are equal:
\begin{equation}
    \int_{0}^\infty \frac{d\omega}{2 \pi} \left[ S_{xx}(\mathbf{k},\omega) - S_{yy}(\mathbf{k},\omega) \right] = 0.
\end{equation}


\section{\label{sec:conclusion} Conclusions}

To summarize, we have shown that:
\begin{enumerate}[label=(\arabic*),wide=0pt,labelindent=\parindent]
    \item The bosonic magnon quasiparticles of a 2D Lifshitz easy-plane antiferromagnet are interacting, but are long-lived and well-defined everywhere in the phase diagram, including in the Lifshitz spin liquid.
    \item Near the Lifshitz point, the spectral weights of processes involving finite numbers of quasiparticles are suppressed by infrared divergent quantum fluctuations.
    \item However, the dynamic structure factor is broadened by contributions from large logarithms (see \eqref{eq:p_n}) which balance the fluctuations.
    \item Precisely at the Lifshitz point, quantum fluctuations force the external probe to excite an infinite number of arbitrarily low energy quasiparticles.
\end{enumerate}
We conclude by noting that the presence of a two-particle continuum is often taken as a signature of magnons fractionalizing into pairs of spinons \cite{Savary2017}. On the other hand the structure factor we have calculated \eqref{eq:skw_xx_yy} is explicitly a two-particle continuum weighted by radiative corrections. However, we have demonstrated in this paper that the quasiparticles of the LSL are not fractional excitations.



\begin{acknowledgments}
We thank Yaroslav Kharkov for useful discussions. This work was supported by the Australian Research Council Centre of Excellence in Future Low Energy Electronics Technologies (CE170100039).
\end{acknowledgments}


\appendix

\section{Mapping between lattice and field theory models \label{app:parameters}}

By using spin wave theory to calculate the magnon dispersion from the $J_1$-$J_3$ model and comparing with the field theory dispersion \eqref{eq:dispersion}, it is possible to more accurately determine the field theory parameters in terms of the microscopic model by ensuring that the two methods predict the same quasiparticle energy. However, it is well known that the ultraviolet physics of the specific lattice model can have important effects which are not captured by spin wave theory. Therefore, the following calculation remains only an approximation.

Consider the Hamiltonian \eqref{eq:j1j3}. The equations of motion for the $\hat{\mathbf{y}}$ and $\hat{\mathbf{z}}$ components of spin are
\begin{subequations}
\begin{align}
    \frac{d}{d t} S_n^{(y)} &= -J_1 \sum_{\langle \ell,n \rangle} S_\ell^{(x)} S_n^{(z)} -J_3 \sum_{\langle\langle\langle m,n \rangle\rangle\rangle} S_m^{(x)} S_n^{(z)}, \\
    \frac{d}{d t} S_n^{(z)} &= J_1 \sum_{\langle \ell,n \rangle} \left( S_\ell^{(x)} S_n^{(y)} - S_\ell^{(y)} S_n^{(x)}  \right) \nonumber \\
    &\qquad + J_3 \sum_{\langle\langle\langle m,n \rangle\rangle\rangle} \left( S_m^{(x)} S_n^{(y)} - S_m^{(y)} S_n^{(x)}  \right).
\end{align}
\end{subequations}
We fix the $\hat{\mathbf{x}}$ axis as the direction of spontaneous staggered magnetization, and then expand around the classical N\'eel state by setting $S^{(x)} = + S$ everywhere on the ``up'' sublattice and $S^{(x)} = - S$ everywhere on the ``down'' sublattice. To leading order in $1/S$, the equations of motion are then linearized:
\begin{subequations}
\begin{align}
    \frac{d}{d t} S_n^{(y)} &= \pm 4 S (J_1 - J_3) S_n^{(z)} , \\
    \frac{d}{d t} S_n^{(z)} &= \mp J_1 S \sum_{\langle \ell,n \rangle} \left( S_n^{(y)} + S_\ell^{(y)} \right) \nonumber \\
    &\qquad \pm J_3 S \sum_{\langle\langle\langle m,n \rangle\rangle\rangle} \left( S_n^{(y)} - S_m^{(y)} \right),
\end{align}
\end{subequations}
taking the upper (lower) sign if site $n$ belongs to the up (down) sublattice. The system can be further simplified by switching to Fourier space $S_n^{(\alpha)} \rightarrow S_{\mathbf{k}}^{(\alpha)} e^{i \mathbf{k}\cdot\mathbf{r}_n}$:
\begin{subequations} \label{eq:spin_momentum_system}
\begin{align}
    \frac{d}{d t} S_{\mathbf{k}}^{(y)} &= \pm 4 S (J_1 - J_3) S_{\mathbf{k}}^{(z)} , \\
    \frac{d}{d t} S_{\mathbf{k}}^{(z)} &= \mp 4 J_1 S \left( S_{\mathbf{k}}^{(y)} + \gamma_{\mathbf{k}} \tilde{S}_{\mathbf{k}}^{(y)} \right) \nonumber \\
    &\qquad \pm 4 J_3 S (1 - \gamma_{2\mathbf{k}}) S_{\mathbf{k}}^{(y)} ,
\end{align}
\end{subequations}
where the tilde denotes a momentum mode belonging to the opposite sublattice, and
\begin{equation}
    \gamma_{\mathbf{k}} = \frac{1}{2}(\cos k_x + \cos k_y ),
\end{equation}
in units of the lattice spacing $a = 1$. The equations \eqref{eq:spin_momentum_system} actually define a system of four coupled linear differential equations: one equation for each $\hat{\mathbf{y}}$ and $\hat{\mathbf{z}}$ component on each sublattice. Diagonalizing the system gives two magnon frequency branches
\begin{equation}
    \omega^2_{\mathbf{k}} = 16 J_1^2 S^2 \left(1 - \frac{J_3}{J_1} \right) \left[ 1 \pm \gamma_{\mathbf{k}} - \frac{J_3}{J_1}(1 - \gamma_{2\mathbf{k}}) \right].
\end{equation}
The lower branch corresponds to the negative sign, and expanding in powers of momentum gives
\begin{align}
    \omega^2_{\mathbf{k}} &\simeq 4 S^2 (J_1 - J_3) \left(J_1 - 4 J_3 \right) k^2 \nonumber \\
    &\qquad + \frac{1}{3} S^2 \left(J_1 - J_3 \right)\left( 16 J_3 - J_1 \right)(k_x^4 + k_y^4).
\end{align}
We first fix the magnetic susceptibility $\chi_\perp = 1/4 J_1$ to agree with the non-frustrated model, and then compare with the field theory dispersion \eqref{eq:dispersion} to find $b_2 = 0$ and
\begin{subequations} \label{eq:params}
\begin{align}
    \rho &= J_1 S^2 \left[ 1 - 5 \left(\frac{J_3}{J_1}\right) + 4 \left(\frac{J_3}{J_1}\right)^2 \right],  \\
    b_1 &= J_1 S^2 \left[ -\frac{1}{12} + \frac{17}{12} \left(\frac{J_3}{J_1}\right) - \frac{4}{3} \left(\frac{J_3}{J_1}\right)^2 \right] .
\end{align}
\end{subequations}
These parameters are consistent with the classical LP $J_3 = J_1/4$, at which $\rho = 0$ and $b_1 = 3 J_1 S^2 /16$. We also note that at $k_x, k_y \approx 0$, the lower branch has an eigenvector with $S^{(z)} = 0$ and the $S^{(y)}$ components of each sublattice oscillating with a  $\pi$ phase difference. This is precisely an excitation of the angle $\varphi$ in the continuum theory.


\bibliography{main}

\end{document}